\documentclass[journal,transmag]{IEEEtran}

\usepackage{epsfig}
\usepackage{cite}

\begin{document}
%

\title{Experimental Demonstration of a Josephson Magnetic Memory Cell with a Programmable $\pi$-Junction} 



\author{\IEEEauthorblockN{Ian M. Dayton \IEEEauthorrefmark{1},
Tessandra Sage \IEEEauthorrefmark{1}, 
Eric C. Gingrich \IEEEauthorrefmark{1}, 
Melissa G. Loving \IEEEauthorrefmark{1}, 
Thomas F. Ambrose \IEEEauthorrefmark{1},\\
Nathan P. Siwak \IEEEauthorrefmark{1},
Shawn Keebaugh \IEEEauthorrefmark{1},
Christopher Kirby \IEEEauthorrefmark{1},
Donald L. Miller \IEEEauthorrefmark{1},\\
Anna Y. Herr \IEEEauthorrefmark{1},
Quentin P. Herr \IEEEauthorrefmark{1},
Ofer Naaman \IEEEauthorrefmark{1}}
\IEEEauthorblockA{\IEEEauthorrefmark{1}Northrop Grumman Systems Corp., Baltimore, Maryland 21240, USA}
\thanks{Corresponding author: A.Y. Herr (email: anna.herr@ngc.com)}}

\markboth{Subject Area: Superconducting Spintronics}%
         {Dayton \MakeLowercase{\textit{et al.}}:Experimental Demonstration of a Josephson Magnetic Memory Cell with a Programmable $\pi$-Junction}
         
%

\IEEEtitleabstractindextext{%
  \begin{abstract}
We experimentally demonstrate the operation of a Josephson magnetic random access memory unit cell, built with a Ni$_{80}$Fe$_{20}$/Cu/Ni pseudo spin-valve Josephson junction with Nb electrodes and an integrated readout SQUID in a fully planarized Nb fabrication process. We show that the parallel and anti-parallel memory states of the spin-valve can be mapped onto a junction equilibrium phase of either zero or $\pi$ by appropriate choice of the ferromagnet thicknesses, and that the magnetic Josephson junction can be written to either a zero-junction or $\pi$-junction state by application of write fields of approximately 5 mT. This work represents a first step towards a scalable, dense, and power-efficient cryogenic memory for superconducting high-performance digital computing.
\end{abstract}

}

\maketitle


\section{Introduction}
Ultra-low power superconducting digital technologies such as Reciprocal Quantum Logic (RQL) \cite{Herr11}, offer to address the energy dissipation challenge that is now facing traditional machines based on complementary metal-oxide semiconductor (CMOS). Even after including the energy cost associated with their cryogenic operation, superconducting logic technologies \cite{Mukhanov11,Holmes13} dissipate 10-100 times less `wall power' than CMOS at comparable clock rates, and are projected to be capable of meeting the U.S.\ Department of Energy exa-scale power dissipation target for high-performance computing \cite{Holmes13,Manheimer15,Holmes15}. While there has been considerable progress in superconductor fabrication technology \cite{Tolpygo16,Johnson10}, automated design tools \cite{Xu17}, and circuit complexity \cite{Herr13,Herr15}, the lack of a suitable memory solution has, hitherto, remained a major risk to the eventual integration of superconducting logic into high-performance computing systems \cite{Holmes13}. Recent efforts to advance the cryogenic memory state-of-the-art beyond superconducting quantum interference device (SQUID)-based memories \cite{Tahara91} include hybrid Josephson-CMOS \cite{VanDuzer13} and magnetic \cite{Herr12,Ohki17} memory solutions. Here, we report on an experimental demonstration of a cryogenic magnetic memory unit cell built in a superconducting integrated circuit, paving the way to a memory solution that is dense, fast, robust, energy-efficient, and compatible with superconducting logic fabrication and signal levels.

Our memory architecture, which we call Josephson magnetic random-access memory (JMRAM) \cite{Herr12}, is based on a magnetic pseudo spin-valve with superconducting electrodes forming a magnetic Josephson junction. Like conventional field-switched MRAM \cite{Slaughter09,Engel05}, JMRAM encodes information in the relative orientation of the two magnetic layers of a spin-valve, and is written by applying bit- and word- write magnetic fields to a selected memory address to set the magnetization of its free layer. Unlike MRAM, readout is not based on sensing the resistance of the bit\textemdash rather, JMRAM readout is anchored in the well-established physics of the Josephson effect in magnetic $\pi$-junctions \cite{Buzdin05,Eschrig11,Eschrig15,Feofanov10}. The magnetic state of a bit is mapped, by appropriate choice of magnetic layer thicknesses, into one of two possible junction ground states differing in their equilibrium superconducting phase. The phase of the magnetic Josephson junction (MJJ), written to zero or $\pi$ depending on the encoded bit, is then accessible for fast, efficient, and high-fidelity readout by a dc-SQUID. Unlike readout schemes that are based on the magnitude of the bit's resistance or critical current and are inherently analog, the JMRAM readout of a phase that can only be zero or $\pi$, is fundamentally digital.

\begin{figure}
\includegraphics[width=\columnwidth]{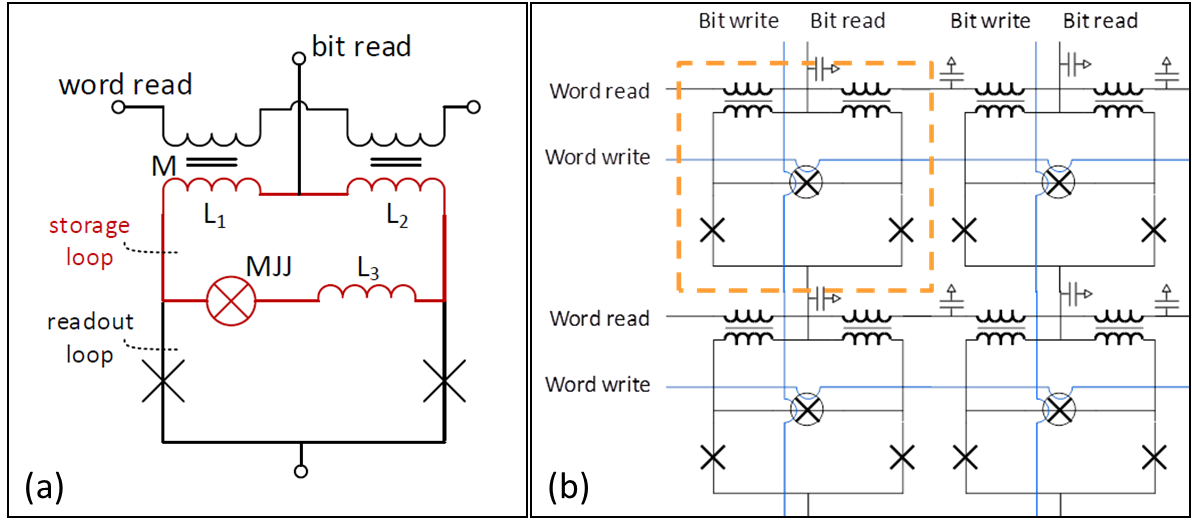}
\caption{\label{fig1} (a) Schematic of the JMRAM unit cell, including an rf-SQUID storage loop (red) that contains the magnetic Josephson junction and a dc-SQUID readout loop. The word-read current applies flux to a selected memory cell such that the bit-read current exceeds the readout SQUID critical current in one memory state, but not in the other. (b) Unit cells can be tiled to form an x-y addressable memory array, with word- and bit-write lines applying hard- and easy-axis fields. Read lines use the cell inductance to form $LC$ ladder transmission lines for fast signal propagation.}
\end{figure}

\section{Experiment}
The JMRAM unit cell is shown schematically in Figure \ref{fig1}(a). An rf-SQUID containing the magnetic junction, labeled as MJJ in the figure, and inductors $L_1$, $L_2$, and $L_3$, forms a storage loop that encloses a spontaneously generated flux $\Phi_0/2$ when the MJJ is in a $\pi$-junction state and zero flux otherwise. To enable storage, the loop linear inductance must be greater than the Josephson inductance of the MJJ. A portion of the stored flux is coupled into the dc-SQUID readout loop and can be sensed by passing a bit-read current through the device. A word-read current tunes the dc-SQUID flux such that in the memory state associated with an MJJ $\pi$ phase ($\Phi_0/2$ in the storage loop), the bit-read current causes the dc-SQUID to switch to its voltage state. In the other memory state, which is associated with an MJJ zero phase, the dc-SQUID critical current is higher than the bit-read current and the SQUID remains in the zero-voltage state. Figure \ref{fig1}(b) shows how cells can be tiled to form an x-y addressable memory array \cite{Herr12}. Fast transmission of read signals through the array is enabled by harnessing the inductance of the cells to form $LC$ transmission lines. The energy dissipated in a write operation on an N-bit word, $E_\mathrm{ww}$ can be estimated as $E_\mathrm{ww}=\eta^{-1}N_b\left(N_wL_bI_b^2+L_wI_w^2\right)/2$ where $N_{b(w)}$ is the size of the array in the bit (word) dimension, $L_{b(w)}$ is the bit (word) write line inductance per cell, $I_{b(w)}$ is a mA-level bit (word) write current, and $\eta$ is the driver efficiency factor\textemdash up to 20\% for RQL drivers \cite{Herr15pump} 

Our magnetic spin-valve Josephson junctions are built with Ni$_{80}$Fe$_{20}$ (NiFe) as a free layer, Ni as a fixed layer, and a Cu spacer that separates the two layers. Additional Cu layers separate the magnetic stack from the Nb electrodes. This particular choice of material combination follows Refs. \cite{Baek14,Gingrich16,MSU17}, which have shown that similar stacks can have substantial critical current densities. In our cell design, the critical current of the MJJ must be higher than that of the readout SQUID junctions; this requirement, together with the desire to size the junction to contain a single magnetic domain, drive the need for high critical current densities. With higher MJJ critical current, the storage loop inductance can be made smaller, resulting in higher memory density and favorable power budget associated with lower write-line inductance per bit.    

The supercurrent $I_{cM}$ through a junction with ferromagnetic (FM) metal barrier decays exponentially with increasing barrier thickness $d$ on a length scale that is inversely proportional to the exchange energy of the FM \cite{Buzdin05}. Superimposed on this decay are oscillations in the critical current $I_{cM}(d)$. These oscillations reverse the sign of $I_{cM}$, and the minimum energy of the junction periodically alternates to favor zero- or $\pi$-phase as a function of thickness on a characteristic scale $\xi_F$. Mapping of the magnetic parallel (P) or anti-parallel (AP) state of a spin-valve with $d_{1,2}$ and $\xi_{F1,2}$ (subscript 1 and 2 refers to first and second magnetic layer) onto a junction phase requires that the sum of the superconducting phases accumulated in the P state $(d_1/\xi_{F1})+(d_2/\xi_{F2})$, and their difference $(d_1/\xi_{F1})-(d_2/\xi_{F2})$ in the AP state, put the junction on opposite-sign lobes of the $I_{cM}(d)$ curve \cite{Crouzy07}. Our approach builds on the work of Ref.\ \cite{Gingrich16}, which clearly demonstrated that a properly tuned Josephson spin-valve functions as a programmable $\pi$-junction whose phase is controlled by the P or AP alignment of the FM layers.   

\begin{figure}
\includegraphics[width=3.3in]{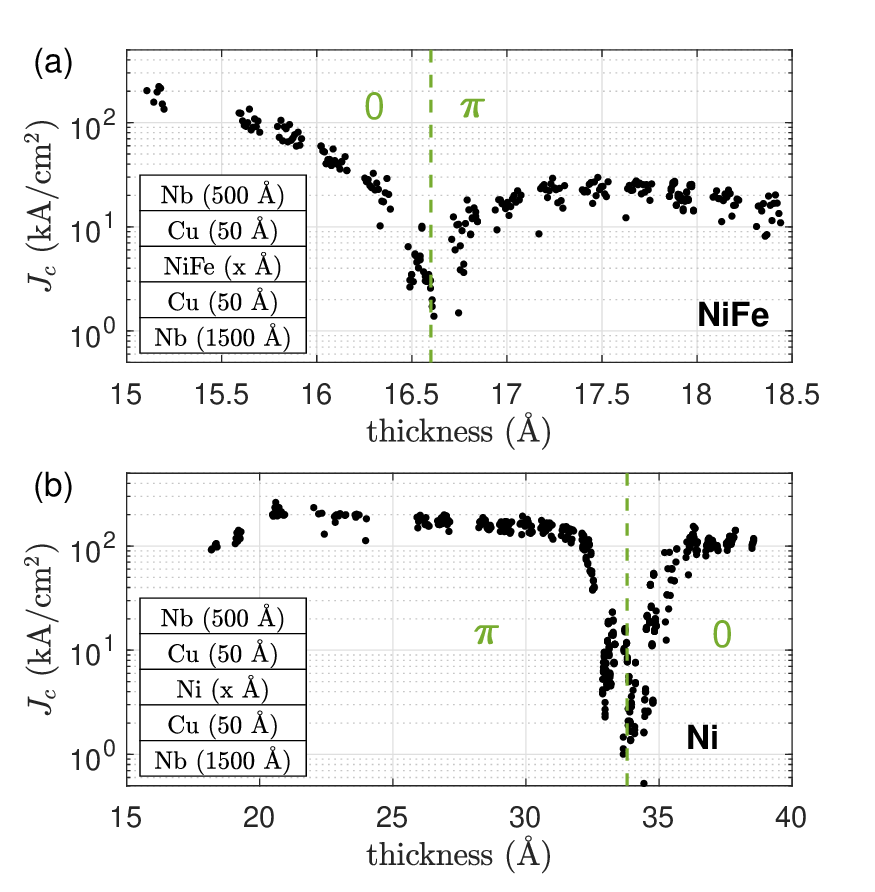}
\caption{\label{fig2} Critical current density $J_c$ of (a) Nb/Cu/NiFe/Cu/Nb and (b) Nb/Cu/Ni/Cu/Nb Josephson junctions vs magnetic barrier thickness. All data in each of the panels were measured on a single wafer, and the thickness gradient was produced by depositing the respective ferromagnetic metal at a 30 degree angle to the wafer normal. Regions of equilibrium 0- or $\pi$-phase are labeled according to Refs.~\cite{Glick17} and \cite{Baek17}. }
\end{figure}

To tune the layer thicknesses in our spin-valves, we first characterize the critical current through each of the FM layers separately as a function of their thickness. We fabricated MJJs containing Cu/NiFe/Cu and Cu/Ni/Cu barriers by depositing the FM layer at a 30 degree angle from the wafer normal, producing a gradient in the FM layer thickness across the wafer. Figure \ref{fig2} shows the critical current density $J_c$, measured at 4.2 K and zero field, of a series of $1~\mu\textrm{m}\times 2~\mu\textrm{m}$ ellipse-shaped MJJs as a function of NiFe [Figure \ref{fig2}(a)] and Ni [Figure \ref{fig2}(b)] thickness across a 150 mm wafer. We see the expected oscillatory decay in the critical current, with Ni having a much longer decay length than NiFe; the range of the data, however, is not sufficient for a meaningful fit to theory. The pronounced dips in $J_c$, at 16.6 {\AA} for NiFe and at 34 {\AA} for Ni, dashed lines in Fig.~\ref{fig2}, indicate these respective thicknesses as $0-\pi$ transitions in these layers, in agreement with Refs. \cite{Glick17} and \cite{Baek17}. With this information, we set the Ni layer thickness in our spin-valve to 33 {\AA}, right at its $0-\pi$ transition thickness, and that of NiFe at 16 {\AA} \textemdash just below the transition. This choice ensures robust mapping between the junction magnetic state and its phase.

\begin{figure}
\includegraphics[width=3.0in]{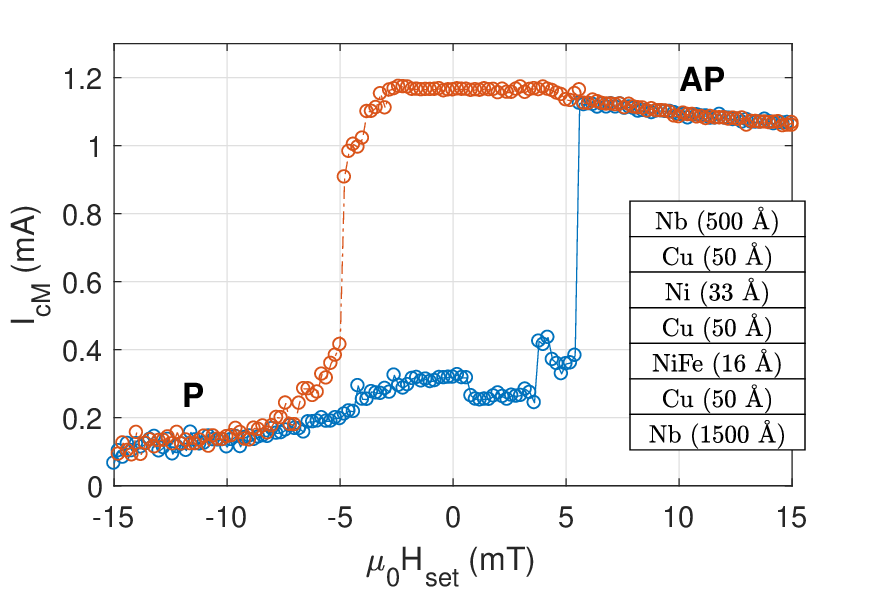}
\caption{\label{fig3} Critical current of a $1~\mu$m$\times2~\mu$m elliptical junction containing the layer stack shown in the inset. Measurements were performed at zero-field following the application of a set field shown on the x-axis. Blue (red) data points represent an increasing (decreasing) $H_{set}$ sweep. The data trace the minor magnetization loop of the junction NiFe free layer.}
\end{figure}

Unit cell devices were fabricated at Northrop Grumman on 150 mm wafers in a fully planarized Nb process with four metal layers, 0.26 $\mu$m minimum linewidth and spacing, featuring both Nb/Al/AlO$_x$/Nb trilayer Josephson junctions and the magnetic spin-valve Josephson junctions shown in the inset of Figure \ref{fig3}. The spin-valve stack was deposited using conventional DC magnetron sputtering from single targets at ambient temperature and a base pressure $<3\times10^{-9}$ torr. In all data shown below, the MJJs are $1~\mu$m $\times 2~\mu$m ellipses, magnetic fields are applied in-plane along the MJJ's easy axis from an off-chip superconducting solenoid, and all measurements were done at 4.2 K.

To probe the critical current of the MJJ, which is not directly accessible in the unit cell, and its dependence on the spin-valve magnetic state, we measure co-fabricated stand-alone MJJ test structures. Representative data are shown in Figure \ref{fig3}. The junction is first initialized to a P state by applying a field of -125 mT, and then we perform a remnant magnetization experiment, in which each of the data points in the figure is acquired at zero field after the application of a set field as shown on the x-axis. This way we can trace the minor magnetization loop of the free layer in our junction. The data clearly shows a switch in the magnetization of the free layer when the set-field is increased to about $\mu_0H=5.5$ mT, associated with a change in the junction critical current from approximately 0.14 mA at -10 mT to 1.1 mA at 10 mT. As the set-field is traced back down through zero, the junction remains in the AP state until $\mu_0H=-4.9$ mT, where it switches back to its initial P state. The hysteretic switching and the existence of two stable states at zero field are the hallmarks of a memory element. In all junctions, we observe $I_{cM,P}<I_{cM,AP}$ because in the P state the net magnetization field of the FM barrier suppresses the zero-field critical current via the Fraunhofer effect \cite{MSU17}, while in the AP state the net field through the barrier, along with the Fraunhofer pattern shift, is significantly lower. This effect becomes much less significant in sub-micrometer junctions \cite{MSU17}. We also observe that while the switching of the free layer often progresses through several intermediate states, the application of $\pm10$ mT saturates the free layer and reliably selects one of the stable states.

\begin{figure}
\includegraphics[width=3.0in]{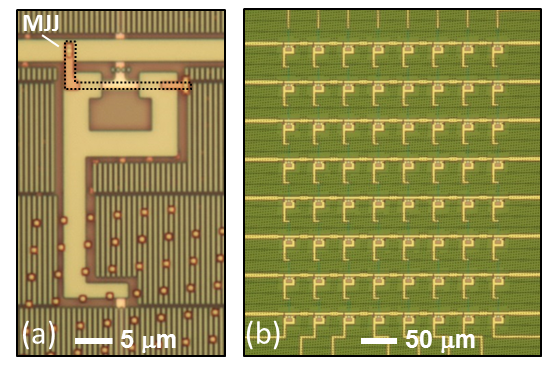}
\caption{\label{fig4} (a) Optical micrograph of the unit cell and (b) an 8x8 array, prior to deposition of the final metal layer. The position of inductor $L_3$, which closes the storage loop in the top layer, is indicated with a dashed line in (a).}
\end{figure}

Having characterized the MJJ both electrically and magnetically, we now turn to the measurement of the unit cell device, whose schematic is shown in Figure \ref{fig1}(a). Fig.~\ref{fig4} shows an optical micrograph of a unit cell device (a) similar to the one measured, and an 8x8 memory array (b). The present design shown in the figure makes no attempt to exercise the full process capability, with lines and spaces many times larger than the minimum design rule. In our device, the inductances of the storage loop were designed to be $L_1=L_2=5.88$ pH, and $L_3=10.93$ pH, and the total critical current of the readout SQUID is 13 $\mu$A. The MJJ critical current in either state of the bit, as can be seen from Figure \ref{fig3}, is such that the MJJ Josephson inductance is considerably smaller than the loop linear inductance, making the storage loop multistable as it can store a number of flux quanta. As we describe below, this is a nuisance that is particular to the current measurement and is not a fundamental property of the technology, nor does it affect our ability to distinguish between the memory states.

When a flux $\Phi_s$ is enclosed in the storage loop, a portion of that flux is mutually induced in the readout loop $\Phi_{r}$ by the current flowing through inductor $L_3$: $\Phi_{r}=\Phi_s\frac{L_3}{L_1+L_2+L_3}\sim0.5\Phi_s$. The ratio $\Phi_{r}/\Phi_s$ was measured in a separate experiment by comparing the readout SQUID modulation period in two cells, where in one of the cells the branch containing $L_3$ and the MJJ was disconnected from the circuit, but were otherwise identical. We observed that severing the MJJ branch reduces the SQUID modulation period by a factor of 2 as expected. Therefore, each $\Phi_0$ change in the flux state of the storage loop will cause the readout SQUID modulation curve $I_{c,sq}(\Phi_{r})$ to shift by half a period. However, a $\pi$-phase shift in the MJJ that is the contrast between memory states will correspond to a change in the stored flux by $\Phi_0/2$, and will offset $I_{c,sq}(\Phi_{r})$ by a quarter of a period. Therefore a zero- to $\pi$-junction transition can be easily distinguished from $\Phi_0$ flux jumps in the large-inductance rf-SQUID storage loop.

\begin{figure}
\includegraphics[width=3.3in]{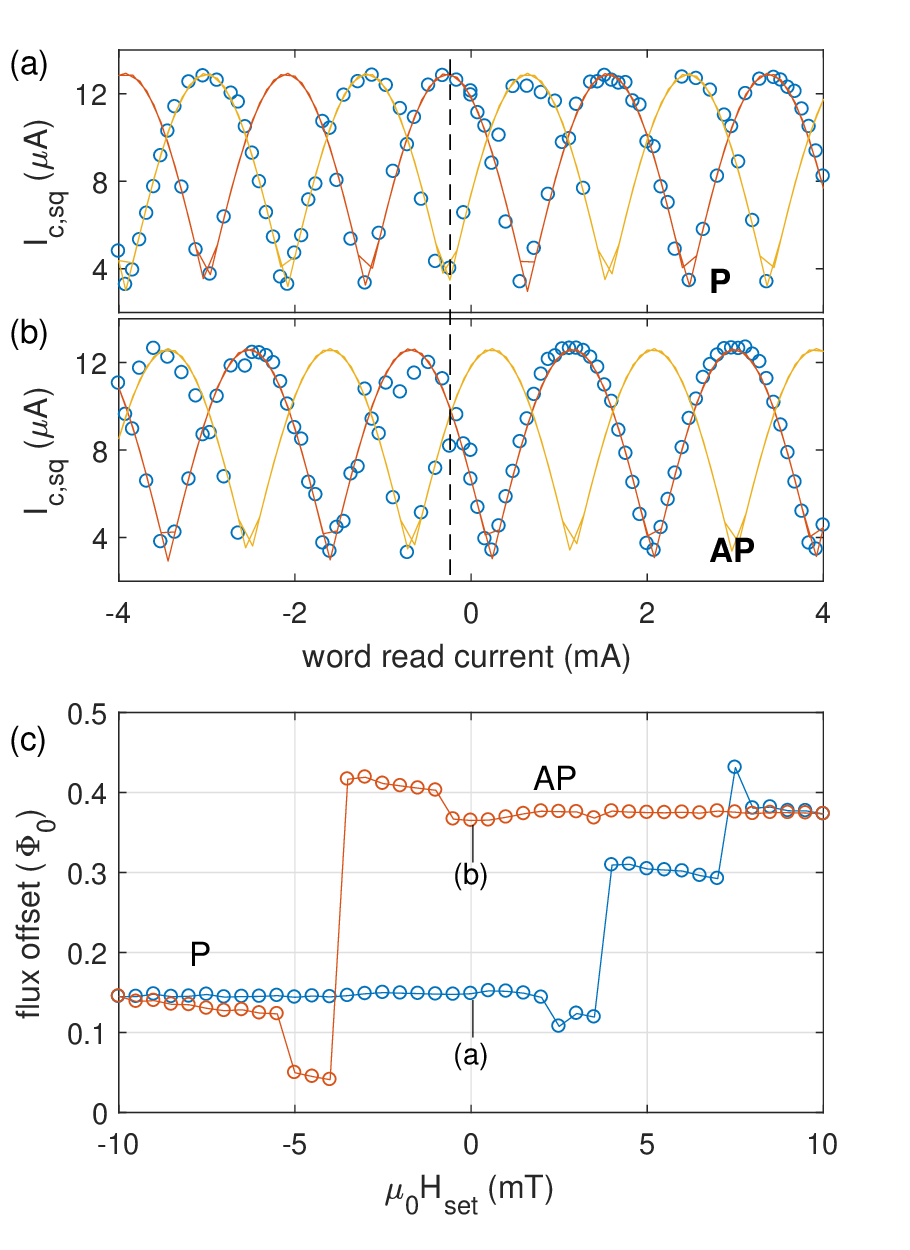}
\caption{\label{fig5} Measurement of the readout SQUID critical current $I_{c,sq}$ vs word-read current, for a parallel (a) and anti-parallel (b) alignment of the spin-valve layers. The quarter-period shift between the (a) and (b) data (see dashed line as a guide) is indicative of a $\pi$ phase change in the MJJ. The $I_{c,sq}$ data (circles) are fit simultaneously to two SQUID modulation curves offset by half a period (red and yellow solid lines). The overall flux offset is extracted from these fits, and is plotted (modulo $\Phi_0/2$) vs set field $H_{set}$ in (c), which traces the free layer minor magnetization loop with blue and red data representing increasing and decreasing fields, respectively. The data points in (c) that were extracted from traces (a) and (b) are marked in the figure.}
\end{figure}

After initializing the cell to a parallel state in a -100 mT field, we apply a set-field pulse $H_{set}$ and then measure the readout SQUID critical current modulation vs word-read current at zero field. The measurement repeats for varying $\mu_0H_{set}$ from -10 mT to 10 mT and back, tracing the free layer minor magnetization loop. Figure \ref{fig5}(a) shows the SQUID critical current $I_{c,sq}(\Phi_{r})$ for $H_{set}=0$ in the parallel memory state of the bit. Each $I_{c,sq}$ data point represent the average over 50 switching events from bit-read current ramps \cite{Fulton74}. As shown in the figure, the $I_{c,sq}$ data traces two modulation curves, offset by half a period, that represent states of the storage loop that differ by integer number of flux quanta. We fit the data simultaneously to the two offset modulation curves (solid lines in the figure), to obtain the overall flux offset of the pattern. Figure \ref{fig5}(b) shows the results of the same experiment but with the bit in the anti-parallel memory state. A comparison of Figures \ref{fig5}(a) and \ref{fig5}(b) clearly shows a quarter-period shift in the SQUID modulation pattern between the P and AP state, indicating a $\pi$ phase shift in the MJJ. Stray field from the junction magnetic barrier can offset the SQUID modulation curve, but we have measured that contribution to amount to less than 0.1 $\Phi_0$, and the observed shifts cannot be explained by that mechanism alone. Note that the experiment is only sensitive to changes in the MJJ phase, and cannot determine whether the $\pi$ (0) phase occurs in the parallel (anti-parallel) state or vice versa.

An important observation from the data of Figures \ref{fig5}(a) and \ref{fig5}(b) is that the magnitude of the MJJ critical current, which changes significantly between the P and AP states (Figure \ref{fig3}), has little to no effect on the readout characteristics. This is a desirable design feature of the JMRAM unit cell because the MJJ critical current depends exponentially on the thickness of a deposited barrier, which is difficult to control to sub-angstrom accuracy across the wafer. The cell's insensitivity to that parameter means that this architecture can be scaled without loss of operating margins due to uncertainties in the barrier thickness. 

Figure \ref{fig5}(c) shows the readout SQUID modulation curve flux offset, modulo $\Phi_0/2$, extracted from measurements like those in panels (a) and (b), as a function of $H_{set}$. The blue and red data points represent, respectively, increasing and decreasing $H_{set}$ sweeps. The data shows that as $H_{set}$ increases, the measured flux offset switches by $\Phi_0/4$ at around 5 mT, corresponding to a switch in the magnetic state of the bit from P to AP alignment. When $H_{set}$ is decreased from 10 mT, the flux offset persists in its AP value past zero field, until the the bit switches back to a parallel state at around -4 mT. In both cases, the switching progresses through a couple of intermediate states that change the flux offset by up to 0.1 $\Phi_0$, consistent with changes in the stray fields picked up by the readout loop. To test the robustness of the saturated P and AP memory state, we performed 300 write/read cycles in which we alternated $\mu_0H_{set}$ between $\pm10$ mT and read the correct phase state with zero errors. Figure \ref{fig5} is the main result of this Letter, and represents an unambiguous demonstration of a programmable Josephson zero-junction to $\pi$-junction transition, written by an applied magnetic field, and read-out by an integrated SQUID.

\section{Conclusion}
We have demonstrated experimentally the operation of a JMRAM memory unit cell, embedding a NiFe/Cu/Ni pseudo spin-valve Josephson junction, and built in a superconducting integrated circuit. We have shown that the magnetization state of the spin-valve, which stores the bit's information, can be mapped onto the junctions's superconducting phase enabling high-fidelity readout by an integrated dc-SQUID. While improvements to the magnetic performance of both free and fixed layers are still required before large memory arrays can be implemented, our experiment lays the basis for this technology, and makes a first step towards a scalable cryogenic memory solution to support power-efficient post-CMOS computing.

\section*{Acknowledgement}
We thank N.O.\ Birge, N.D.\ Rizzo, and N.\ Newman for technical assistance and discussions, R. Pownall, J. Robinson, K. Holman, K. Dinh, and A. Sidorov for measurements support, and S. Van Campen, S. Weiss, and J. Stanbro for program support. We thank W.H. Rippard, M.L. Schneider, and P. Hopkins for independent verification of our experimental results. This work was supported by a Northrop Grumman IRAD program and by the Office of the Director of National Intelligence (ODNI), Intelligence Advanced Research Projects Activity (IARPA), via U.S. Army Research Office contract W911NF-14-C-0115. The views and conclusions contained herein are those of the authors and should not be interpreted as necessarily representing the official policies or endorsements, either expressed or implied, of the ODNI, IARPA, or the U.S. Government.

\bibliographystyle{IEEEtran}
\bibliography{unit_cell_transmag}

\end{document}